
\documentstyle{amsppt}
\TagsOnRight
\catcode`\@=11
\def\logo@{}
\catcode`\@=13
\parindent=8 mm
\magnification 1200
\hsize = 6. true in
\vsize = 8.5 true in
\hoffset = .4 true in
\parskip=\medskipamount
\baselineskip=14pt
\redefine\cite#1{{\bf[#1]}}
\def \di{\partial}
\def\rom{\roman}

\def \smaller {\eightpoint}
\def \wt {\widetilde}

\def \mt {\mapsto}
\def \ra {\rightarrow}

\def \lmt {\longmapsto}
\def \a {\alpha}
\def \b {\beta}
\def \d {\delta}

\def \g {\gamma}

\def \k {\kappa}

\def \l {\lambda}

\def \s {\sigma}

\def \pmu {p_{\xi}}
\def \ss {\subset}
\def \lg {\widetilde{\frak g}}

\def \lgp {\widetilde{\frak g}_+}
\def \lgm {\widetilde{\frak g}_-}
\def \lgmp {\widetilde{\frak g}_{\mp}}
\def \lgpm {\widetilde{\frak g}_{\pm}}

\def \CC {\Cal C}
\def \DD {\Cal D}

\def \HH {\Cal H}
\def \II {\Cal I}

\def \NN {\Cal N}

\def \di {\partial}

\def \ln{\text{ln}}
\def \tr{\text{tr}}
\def \Ad{\text{Ad}}
\def \res{\text{res }}

\def \prh{p_\rho}
\def \pmu{p_\mu}
\def \Arh{A_\rho}
\def \Amu{A_\mu}
\phantom{0}\vskip -.5 true in
\noindent hep-th/9406077 \hfill CRM-2889 (1994)
\vskip .5 true in
\topmatter
\title
$R$--Matrix Construction of Electromagnetic Models for the Painlev\'e
Transcendents
\endtitle
\leftheadtext{J\. Harnad and M\. Routhier}
\rightheadtext{ Electromagnetic models for the Painlev\'e Transcendents}
\thanks Research supported in part by the
Natural Sciences and Engineering Research Council of Canada and the Fonds FCAR
du Qu\'ebec.
\endthanks
\endtopmatter
\centerline{\smc J.~Harnad}
\smallskip
{\smaller \centerline{ \it Department of Mathematics and Statistics, Concordia
University }
\centerline{{\it  7141 Sherbrooke W., Montr\'eal, Canada H4B 1R6,} \
and }
\centerline{ \it  Centre de recherches math\'ematiques,
Universit\'e de Montr\'eal }
\centerline{ \it  C.~P. 6128, Succ. centre--ville, Montr\'eal, Canada H3C 3J7}
\centerline{e-mail:
{\it harnad\@alcor.concordia.ca} or {\it harnad\@mathcn.umontreal.ca}}}
\medskip
 \centerline{\smc M.~Routhier}
\smallskip
{\smaller \centerline{ \it Department of Mathematics and Statistics,
Concordia University}
\centerline {\it 7141 Sherbrooke W., Montr\'eal, Canada H4B 1R6}
\centerline{e-mail: {\it manon\@neumann.concordia.ca}}}
\bigskip \bigskip
\centerline{\bf Abstract}
\bigskip
\baselineskip=10pt
\centerline{
\vbox{
\hsize= 5.5 truein
{\smaller
The Painlev\'e transcendents $P_{\rom{I}}$--$P_{\rom{V}}$ and their
representations as isomonodromic deformation equations are derived as
nonautonomous Hamiltonian systems from the classical $R$--matrix
Poisson bracket structure on the dual space $\wt{\frak{sl}}_R^*(2)$ of the loop
algebra  $\wt{\frak{sl}}_R(2)$. The Hamiltonians are obtained by
composing elements of the Poisson commuting ring of spectral invariant
functions on $\wt{\frak{sl}}_R^*(2)$ with a time--dependent family of Poisson
maps whose images are $4$--dimensional rational coadjoint orbits in
$\wt{\frak{sl}}_R^*(2)$.   Each system may be interpreted as describing a
particle moving on a surface of zero curvature in the presence of a
time--varying
electromagnetic field. The Painlev\'e equations follow from reduction of these
systems by the Hamiltonian flow generated by a second commuting element in
the ring of spectral invariants.}}}
 \baselineskip=14pt
\bigskip
\subheading{Introduction}
\nobreak

   The Painlev\'e transcendents $P_{\rom{I}}$--$P_{\rom{VI}}$ \cite{I} may be
interpreted as deformation equations preserving the monodromy of first order
matrix differential operators of the form
$$
\DD_\l = {{\di}\over{\di\l}} - \NN(\l),  \tag{0.1}
$$
where $\NN(\l)$ is a rational, matrix--valued function of $\l$ (cf.
\cite{JM}).  It has long been known that they may also be viewed as
nonautonomous Hamiltonian systems (see, e.g., \cite{Ok} and references
therein). In \cite{H, HW2}, it was demonstrated that the relation between the
Hamiltonian structure and the isomonodromic deformation property follows
naturally from a classical $R$--matrix structure on the loop algebras
$\wt{\frak{gl}}(2)$ and $\wt{\frak{gl}}(3)$ . Following the general scheme of
\cite{H} for obtaining isomonodromic deformation equations from spectral
invariants on loop algebras, the Hamiltonians are obtained through equivariant
moment map embeddings of a symplectic vector space ($\bold{R}^4$, $\bold{R}^6$
in the case of the Painlev\'e equations), or some Hamiltonian quotient thereof,
in the dual space $\lg_R^*$ of a loop algebra $\lg_R$, with Lie-Poisson
structure
determined by the ``split'' classical $R$--matrix  \cite{ST, FT}
$$
R={1\over2} (P_+
-P_-).  \tag{0.2}
$$
   (Here $P_{\pm}$ denotes projections to the positive and negative powers in
the loop parameter.) The relevant Hamiltonians are obtained by restricting
elements of the Poisson commuting ring of spectral invariants on $\lg_R^*$
to the image of these Poisson maps. The isomonodromic property follows from a
modification of the isospectral equations induced by such Hamiltonians, taking
into account explicit time dependence of the parameters defining the moment
map.

   In the case of the Painlev\'e transcendents $P_{\rom{I}}$--$P_{\rom{VI}}$,
the resulting Hamiltonians all turn out to have the simple form
$$
\HH ={1\over 2m }\sum_{i,j=1}^n g^{ij} (y_i + A_i
(x))(y_j + A_j (x))  + V(x),  \tag{0.3}
$$
where $n=2$ for $P_{\rom{I}}$--$P_{\rom{V}}$ and $n=3$ for $P_{\rom{VI}}$. Here
$g^{ij}$, $A_i$ and $V$ denote the components of a symmetric
contravariant tensor field, a covector and a scalar field respectively, and
$\{x_i,y_i\}_{i=1 \dots n}$ are Cartesian canonical coordinates on
$\bold{R}^{2n}$.  After suitable reduction under a  $1$--parameter group of
Hamiltonian symmetries, the  equations $P_{\rom{I}}$--$P_{\rom{V}}$ result. (To
obtain $P_{\rom{VI}}$,  reduction under a Hamiltonian $\frak{Sl}(2,R)$ action
is
required). However, in \cite{HW2}, only for the  case $P_{\rom{V}}$ was the
tensor $g^{ij}$ found to be nonsingular, allowing an interpretation of its
inverse as a Riemannian metric and, correspondingly, the fields $A_i$ and $V$
as
magnetic and electric potentials.

  In the present work it will be shown that, by making suitable modifications
of the Hamiltonians $\HH_{\rom{I}}$--$\HH_{\rom{V}}$ of \cite{H, HW2}, obtained
by the addition of terms involving only the invariant generating the
symmetry group,  one can obtain systems in which not only is the tensor
$g^{ij}$  invertible, but the corresponding metric tensor is Euclidean. The
result is a simple, flat--space $2$--dimensional electromagnetic model for each
case, yielding the Painlev\'e  transcendents $P_{\rom{I}}$--$P_{\rom{V}}$ after
symmetry reduction.  For each of the modified systems, this also provides an
isomonodromy representation following from the classical $R$--matrix structure.

   In Section 1, the notation for loop algebras is given, and the
basic results concerning isospectral flow and isomonodromic deformations
following from the classical $R$--matrix structure on loop algebras are
recalled.  In Section 2, for each of the Painlev\'e transcendents
$P_{\rom{I}}$--$P_{\rom{V}}$,  the relevant moment map embedding and spectral
invariant Hamiltonians are given and Hamilton's equations prior to reduction
are
derived in their isomonodromic deformation form.  These Hamiltonians all turn
out to have the simple electromagnetic form (0.2), with $g^{ij}$ the inverse of
a Euclidean metric tensor. These are are then re-expressed in terms of
Cartesian
coordinates for this metric, the vector and scalar potentials are identified,
and
the reduced forms leading to the Painlev\'e equations are derived.
\bigskip
\subheading{1. Poisson Embeddings in Loop Algebras and the Classical
$R$--Matrix}
\medskip \noindent
{\it 1a.\quad Loop algebras and  $R$--matrix structure \hfill}
\nobreak

Let $\frak g$ be a matrix Lie algebra.  (For the purpose of this work, it will
be sufficient to just consider $\frak g = \frak{sl}(2)$.) Define the loop
algebra $\lg$ to be the set of smooth maps $X:\CC \ra {\frak g}$, where $\CC$
is
a circle centered at the origin in the complex  $\l$--plane. We may split $\lg$
as
a vector space direct sum
$$
\lg = \lgp \oplus \lgm,   \tag{1.1}
$$
where $\lgp$ and  $\lgm$ are the subalgebras consisting of elements admitting
holomorphic extensions to the interior and exterior of $\CC$, respectively,
with the latter vanishing at $\l=\infty$.

  The $\Ad$--invariant scalar product
$$
<X,Y> :={1\over 2\pi i}\oint_{\CC} \tr (XY) d\l, \quad X,\ Y \in \lg,
\tag{1.2}
$$
gives an identification of $\lg$  as a dense subspace of its dual space, which
will henceforth be denoted as $\lg^*$. The splitting (1.1) may thus be
interpreted as  $$
 \lg^* = \lgp^* \oplus \lgm^*,  \tag{1.3}
$$
where $\lgpm^*$ is identified with the orthogonal annihilator
$(\lgmp)^{\perp}$ with respect to (1.2).

    Denoting by $P_+$,   $P_-$ the projections of $\lg$ to $\lgp$, $\lgm$,
respectively, we may define a ``classical $R$--matrix'' \cite{ST} as the
endomorphism:
$$
R = {1\over2}(P_+ - P_-).  \tag{1.4}
$$
The bracket
$$
[X,\ Y]_R := [R(X),\ Y] + [X,\ R(Y)]  \tag{1.5}
$$
determines a second Lie algebra structure on $\lg$, which is essentially the
direct sum of the algebras $\lgp$, $\lgm$. Denote by $\lg_R$ the Lie algebra
endowed with this new bracket and $\lg_R^*$ the corresponding dual space. The
associated Lie-Poisson bracket is then
$$
\{f,\ g\}_{_R}(\NN) := <\NN,[df(\NN),dg(\NN)]_R>, \quad \NN\in \lg_R^*
\tag{1.6}
$$
\medskip \noindent
{\it 1b.\quad Isomonodromic deformations and moment map embeddings \hfill}
\nobreak

Using the general method of moment map embeddings in loop algebras
developed in \cite{AHP, AHH, HW1}, parametric families of Poisson
maps
$$
J_A: M \ra \frak g_A \subset \lg^*_R, \tag{1.7}
$$
may be defined on a symplectic vector space $M$ (which, in the
following,  will just be $\bold{R}^4$), taking their values in a Poisson
subspace ${\frak g}_A \ss \lg^*_R$ consisting of elements $\NN$ of the
form
$$
\NN(\l) :=  \sum_{l = 0}^{n_0}
{N_{0,l} \l^{l}} + \sum_{i=1}^n \sum_{a_i = 1}^{l_i}
{{N_{i, l_i}}\over{(\l - \a_i)^{l_i}}}. \tag{1.8}
$$
For the purpose of deriving isomonodromic deformation equations, in which the
$\a_i$'s may be time--dependent, it is best to choose the circle $\CC$
defining the splitting (1.1) ``at $\infty$'', in order that none of the
time--dependent poles traverse $\CC$ at finite $t$; i.e., the poles at
$\l=\a_i$
are taken as interior to $\CC$, while $\l=\infty$ is exterior. The Lie--Poisson
structure (1.6) on $\lg_R$ may be expressed in terms of the matrix elements by
$$
[\NN_{ij}(\l), \NN_{kl}(\mu)] = {\left(\NN_{il}(\l) -
\NN_{il}(\mu)\right)\d_{jk} - \left(\NN_{kj}(\l) -
\NN_{kj}(\mu)\right)\d_{il}\over \l-\mu}, \tag{1.9}
$$
and the maps $J_A$ are Poisson maps with respect to this structure.  For the
case
$\frak{g}=\frak{sl}(2)$, $\NN(\l)$ has the form
$$
\NN(\l)=\pmatrix h(\l) & e(\l) \\ f(\l) & -h(\l)\endpmatrix,  \tag{1.10}
$$
where $e(\l), \ f(\l), \ h(\l)$ are rational functions of $\l$ satisfying the
Poisson bracket relations
$$
\align
\{h(\l), \ e(\mu)\}&= {e(\l) - e(\mu)\over \l-\mu}   \tag{1.11a}\\
\{h(\l), \ f(\mu)\}&= -{f(\l) - f(\mu)\over \l-\mu}   \tag{1.11b}\\
\{e(\l), \ f(\mu)\}&= 2{h(\l) - h(\mu)\over \l-\mu}.    \tag{1.11c}
\endalign
$$

  In \cite{ [H, HW2]} such maps were used, within the classical $R$--matrix
framework, to generate nonautonomous Hamiltonian equations which, upon
reduction
by suitable symmetries, lead to the the six Painlev\'e transcendents. This
approach is based on the following construction.

Let $\II_A$ be the ring of spectral invariants on $\lg^*_R$, restricted to
$\frak g_A$.  The classical $R$--matrix form of the Adler-Kostant-Symes
theorem \cite{ST, HW1} states that:
\item{1.} All the elements in $\II_A $ Poisson commute with respect to the
Lie--Poisson bracket (1.9) (and hence, so do their pullbacks under the
Poisson moment maps $J_A$).
\item{2.} The equations of motion for $\HH \in \II_A $ are given by the
isospectral equation
$$
\dfrac{d\NN}{dt} = [P_{\s}(\d\HH), \NN]  \tag{1.12}
$$
where, for any ${\sigma} \in \bold{R}$, $P_{\sigma}$ is the endomorphism
$$
P_{\s} :={1\over 2} ({\s}+1)P_+ + {1\over 2}(\s -1) P_-, \quad
\tag{1.13}
$$

\noindent
(The fact that $\HH$ is in the ring $\II_A $ of spectral invariants implies
that $\d\HH(\NN)$ commutes with $\NN$, so the various choices of $\s\in
\bold{R}$
in (1.13) give equivalent equations (1.12).)

   If $\NN$ is understood as the image of $J_A$, eq.~(1.12) may be viewed as
a consequence of Hamilton's equations on $M$ corresponding to the Hamiltonian
$\HH\circ J_A$. If the Poisson map $J_A$ depends explicitly on the time
parameter
$t$,  this becomes a nonautonomous Hamiltonian system and eq\. (1.12) must be
modified to take into account the total derivative of $\NN$:
$$
\dfrac{d\NN}{dt} =
[P_{\s}(\d\HH),\ \NN] + \NN_t, \tag{1.14}
$$
where $\NN_t$ denotes derivation with respect to the explicit $t$--dependence.
If,
furthermore, $\HH$ is such that,  for some value of $\s$, we have
$$
{d(P_{\s} (\d\HH))\over d\l} = \NN_t,  \tag{1.15}
$$
then this system is equivalent to the commutation relation
$$
[\DD_t,\ \DD_{\l}]=0,  \tag{1.16}
$$
for the operators
$$
\align
\DD_\l :=& {\di \over\di \l} - \NN,  \tag{1.17a}\\
\DD_t  :=& {\di \over \di t} - P_{\s}(\d\HH),  \tag{1.17b}
\endalign
$$
and hence determines deformations of the operator $\DD_\l$ that preserve the
monodromy about the poles of $\NN(\l)$. More generally, if $\HH$ splits into a
sum of terms
$$
\HH=\sum_{a=1}^l \HH_a, \tag{1.18}
$$
each of which is a spectral invariant, we may replace $P_\s(\d\HH)$ in
eqs.~(1.12)--(1.17a,b) by the sum
$$
\sum_{a=1}^l P_{\s_a} (\d\HH_a)  \tag{1.19}
$$
for distinct values $\{\s_a\}_{a=1,\dots l}$, since each term $\d\HH_a(\NN)$
individually commutes with $\NN$.

   For $\frak{g}=\frak{sl}(r)$, a complete set of generators for the ring
$\II_A$ is given by the functions
$$
h_m:={1\over 2\pi i} \oint_\CC \l^m \tr(\NN^2(\l))d\l, \qquad m \in\bold{Z}.
\tag{1.20}.\
$$
The number of these that are independent on the image $J_A(M)$ depends on the
pole
structure of $\NN(\l)$. When $J_A$ is an immersion, this number will, in
general,
equal ${1\over 2} \dim(M)$ ($=2$ in the examples to follow), which means the
associated autonomous systems (1.12) on $J_A(M)$ are completely integrable on
``generic'' orbits.

\medskip \noindent
{\it 1c.\quad Electromagnetic systems and reductions to the Painlev\'e
equations
\hfill}
\nopagebreak

The Hamiltonians $\HH_A\in \II_A $ used here to obtain the Painlev\'e equations
$P_{\rom{I}}$--$P_{\rom{V}}$  are obtained from those defined in \cite{H, HW2}
by adding an appropriate constant of motion defined in terms of the second
generator $a\in \II_A$ of the ring of invariants. They will all be of the form
$$
\HH_A^* = \HH_A + k a^2,  \tag{1.21}
$$
for $A=I, \dots V$, where $\HH_A$ is the Hamiltonian of \cite{H, HW2} and $k$
is
independent of the phase space variables, but may  be explicitly
$t$--dependent. The main point is that the factor $k$ may always be chosen so
that $\HH_A^*$, is still of the form (0.2), but with the contravariant tensor
$g^{ij}$ invertible, and such that the corresponding covariant tensor $g_{ij}$
is
interpretable as a flat Euclidean metric. Since Hamilton's equations for the
$\HH_A$'s reduce, under the flow generated by the invariant $a$, to the
Painlev\'e
equations, and $k a^2$ just adds a constant to the reduced Hamiltonian, this
does
not alter the reduced equations.

   In each case, we introduce new canonical coordinates $(\rho, \mu, \prh,
\pmu)$,
such that $(\rho, \mu)$ are Cartesian coordinates on the configuration space
with
respect to the modified Euclidean metric and $(\prh, \pmu)$ are their conjugate
momenta. In these coordinates, the Hamiltonians (0.2) take the simple form
$$
\HH_A^*={1\over 2 m}\left( (\prh +\Arh)^2 +
{1\over 2}(\pmu +\Amu)^2\right) +V(\rho, \mu), \tag{1.22}
$$
where the ``mass'' $m$ may be time dependent.
 The equations of motion are therefore of the $2$--dimensional Lorentz form:
$$
\align
(m\dot{\rho})\dot{} &= \phantom{-}B\dot{\mu} +E_\rho  \tag{1.23a}\\
(m\dot{\mu})\dot{}  &= -B\dot{\rho} +E_\mu,  \tag{1.23b}
\endalign
$$
where
$$
B=A_{\rho, \mu} -A_{\mu,\rho}  \tag{1.24}
$$
is the magnetic field and
$$
(E_\rho, E_\mu) = \left(-{\di V\over \di \rho} +\dot{A_\rho},\
 -{\di V\over \di\mu} + \dot{A_\mu}\right) \tag{1.25}
$$
the electric field. The presence of the additional symmetry group generated by
the invariant $a$ reduces these to a single second order ODE, and the
resulting systems give the Painlev\'e equations $P_{\rom{I}}$--$P_{\rom{V}}$.

  In the following section, using the Poisson embeddings of
refs.~\cite{H, HW2}, we derive the explicit form of the Hamiltonians, the
isomonodromic deformation equations and the Lorentz equations of motion
(1.23a,b)
for each of the Painlev\'e transcendents $P_{\rom{I}}$--$P_{\rom{V}}$,
\bigskip
\subheading{2. Spectral invariant Hamiltonians, isomonodromy representation
and electromagnetic systems}
\medskip \nobreak
\noindent
{\it 2a.\quad  Painlev\'e $\rom{I}$ \hfill}

\nobreak
The moment map (1.7) is defined for this case (cf. \cite{HW2}) as
$$
\align
J_I(x_1,x_2,y_1,y_2) =
\NN(\l) :=& \l^2 \left(\matrix 0 & 1\\0 & 0\endmatrix\right)
+ \l \left(\matrix x_1 & x_2\\ \k & -x_1\endmatrix\right) \\
& +  \left(\matrix -y_2 + x_1 x_2 & y_1 + {t/2}\\
 -x_1^2 - \k x_2 & y_2 - x_1 x_2\endmatrix\right), \tag{2.1}
\endalign
$$
where $\{x_1,x_2,y_1,y_2\}$ are Cartesian canonical coordinates on
$\bold{R}^4$, $\k$ is a positive constant and $t$ is the deformation parameter.
It is easily verified that the matrix entries satisfy (1.11a-c), so this is a
Poisson map. The Hamiltonian $\HH^*_{\rom{I}} \in \II_A $ is defined as
$$
\HH^*_{\rom{I}}:= \HH_{\rom{I}} + {a^2\over 2},   \tag{2.1}
$$
where
$$
\align
\HH_{\rom{I}} := &{1\over 4}\res_{\l=0}\tr(\l^{-1}{\NN^2(\l)}) \\
= &{1\over2}\left({(y_2 - x_1x_2)}^2 - (x_1^2 + \k x_2)(y_1 + {t\over2})\right)
\tag{2.2}
\endalign
$$
and the commuting invariant $a\in \II_A $ is
$$
\align
a :=& {1\over2}\res_{\l=0}\tr(\l^{-2}{\NN^2(\l)})\\
= & \k(y_1 - x_2^2) - 2 x_1 y_2 + x_1^2 x_2. \tag{2.3}
\endalign
$$

The differential is given by
$$
\d\HH_{\rom{I}}^*=\d\HH_{\rom{I}} +a\d a  ={1\over 2\l}\NN(\l) +{a\over
\l^2}\NN(\l),  \tag{2.4}
$$
so, in the notation of eqs.~(1.13), we have
$$
P_{1}(\d\HH_{\rom{I}}^*) = {\l\over 2} \pmatrix 0&1\\0&0\endpmatrix
+{1\over 2}\pmatrix x_1 & x_2 \\ \k & -x_1 \endpmatrix+ \pmatrix 0 & a \\ 0 &
0 \endpmatrix.
\tag{2.5}
$$
It follows that eq.~(1.15) is satisfied for $\s=1$, and Hamilton's
equations are equivalent to the isomonodromic deformation equation (1.16),
with $\NN(\l)$ given by (2.1), and
$$
\DD_t  := {\partial{}\over \partial t} -
 {\left(\matrix 0 & {\l \over2} + a\\0 & 0\endmatrix\right)} -
{1\over2} \left(\matrix x_1 & x_2\\ \k & -x_1\endmatrix\right).  \tag{2.6}
$$

   Since $\HH^*_{\rom{I}}$ is given by:
$$
\align
\HH^*_{\rom{I}} =& {1\over2}(\k^2 y_1^2 + (1 + 4 x_1^2)y_2^2 - 4 \k x_1 y_1 y_2
\\
& +(4 \k x_1x_2^2 -2x_1x_2 - 4x_2x_1^3)y_2 +
(2\k x_2 x_1^2 - 2\k^2 x_2^2 - \k x_2 - x_1^2)y_1 \\
& +x_1^2x_2^2 - {t x_1^2\over2}
 - {t\k x_2\over2} + \k^2 x_2^4 - 2\k x_1^2x_2^3 +x_2^2 x_1^4),
\tag{2.7}
\endalign
$$
we may take $m=1$ in eq.~(0.2), and the contravariant
tensor $g^{ij}$ is
$$
g^{ij}=\left(\matrix \k^2 & -2\k x_1\\
 -2\k x_1 & 1+4x_1^2\endmatrix\right). \tag{2.8a}
$$
The inverse
$$
g_{ij} ={1\over \k^2} \left(\matrix 1 +4x_1^2 & 2\k x_1\\ 2\k x_1 &
\k^2\endmatrix\right) \tag{2.8b}
$$
defines a new Euclidean metric, for which Cartesian coordinates may be chosen
as
$$
\rho ={x_1 \over \k},  \quad \mu = x_2 +{x_1^2\over \k}, \tag{2.9a}
$$
with conjugate momenta
$$
\prh = \k y_1 - 2x_1 y_2,  \quad  \pmu = y_2.  \tag{2.9b}
$$
In terms of these, the Hamiltonian is
$$
\align
\HH^*_{\rom{I}}(\mu,\rho,\pmu,\prh)& = {1\over2}(\prh^2 + \pmu^2) -
(2\k\mu\rho - \k^2\rho^3)\pmu \\
&+( 3\k^2\rho^2\mu - 2\k^3\rho^4 - \k\mu^2 - {\mu\over2})\prh +
 {{\k^2\rho^2\mu^2}\over2} - \k^3\mu\rho^4 +{{\k^4\rho^6}\over2}\\
&+
{{\k^2\mu^4}\over2}
 - 3\k^3\mu^3\rho^2 + {{13\k^4\mu^2\rho^4}\over2}
 - 6\k^5\mu\rho^6 + 2\k^6\rho^8 - {{\k t\mu}\over4}. \tag{2.10}
\endalign
$$
The vector and scalar potentials in this case are therefore:
$$
\align
A =&  (3\k^2\rho^2\mu - \k\mu^2 - {\mu\over2} -
2\k^3\rho^4)d\rho +
(\k^2\rho^3 - 2\k\mu\rho)d\mu   \tag{2.11a} \\
V =& -{{\k t\mu}\over 4} - {{\k\mu^3}\over2} -{\mu^2\over 8}, \tag{2.11b}
\endalign
$$
and the constant of motion $a$  is
$$
a = \prh - 2\k^3\rho^4 - \k\mu^2 + 3\k^2\mu^2\rho^2. \tag{2.12}
$$

   A simplification is obtained by applying the gauge transformation
$$
A\mt A+d\l  =-{\mu\over 2}d\rho :=\wt{A}, \tag{2.13}
$$
where
$$
\l:=\k\mu^2 \rho -\k^2\rho^3\mu +{2\over 5}\k^3\rho^5,  \tag{2.14}
$$
which is equivalent to making the canonical coordinate transformation
$$
(\rho, \mu, \prh, \pmu) \lmt (\rho, \mu, \wt{p}_\rho, \wt{p}_\mu),  \tag{2.15}
$$
where
$$
\wt{p}_\rho=\prh -\l_{,\rho}, \quad  \wt{p}_\mu=\pmu -\l_{,\mu}.   \tag{2.16}
$$
In terms of these coordinates, the Hamiltonian (2.10) becomes
$$
\wt{\HH}^*_{\rom{I}}(\mu,\rho,\wt{\rho},\wt{p}_\mu) =
{1\over2}\left(\wt{p}_\rho^2
+ \left(\wt{p}_\mu -{\mu\over 2}\right)^2\right)
-{{\k t\mu}\over 4} - {{\k\mu^3}\over2} -{\mu^2\over 8} \tag{2.17}
$$
and the corresponding invariant is
$$
\wt{a}:= \wt{p}_\rho.  \tag{2.18}
$$
The magnetic field $B$ is just the constant $-{1\over 2}$, and the Lorentz
equations (1.23a,b) become
$$
\align
\ddot{\rho} &=-{1\over 2}\dot{\mu}  \tag{2.19a}\\
\ddot{\mu} &={1\over 2} \dot{\rho} + {3\k\over 2}\mu^2+ {\mu \over 4}
  +{\k t\over 4}.  \tag{2.19b}
\endalign
$$
Eq.~(2.19a) is equivalent to the fact that
$$
 \dot{\rho} +{\mu\over 2} = a \tag{2.20}
$$
is constant. Substituting in (2.19b) then gives
$$
\ddot{\mu} = {3\k\over2}\mu^2 + {\k \over4}t + {a\over 2}, \tag{2.21}
$$
which, for $\k=4$, $a=0$, is the standard form of $P_{\rom{I}}$.

   Equivalently, fixing the value of the invariant $a$ reduces the
Hamiltonian (2.17) to
$$
\wt{\HH}^{red}_I(\mu, \pmu,t) = \wt{p}_\mu^2 -{\mu \pmu\over 2}
-{{\k t\mu}\over 4} - {{\k\mu^3}\over2} +{1\over2}a^2, \tag{2.22}
$$
and Hamiltonian's equations again give eq.~(2.21) after elimination of
the momentum $\pmu$.
\medskip \noindent
{\it 2b.\quad  Painlev\'e $\rom{II}$ \hfill}

\nobreak
The moment map (1.7) for this case is defined as
$$
\align
J_{II}(x_1,x_2,y_1,y_2) =
\NN(\l) := & \l^2 \left(\matrix {{\k}\over2} & 0\\0 &
-{{\k}\over2}\endmatrix\right) + \l \left(\matrix 0 & -\k y_1\\ x_2 &
0\endmatrix\right)  \\  & + \left(\matrix x_2y_1 + {t\over2} & -\k y_2\\ x_1 &
-x_2y_1 - {t\over2}\endmatrix\right), \tag{2.23}
\endalign
$$
where $\k$ is a nonzero constant and $t$ is the deformation parameter. Again,
the Poisson bracket relations (1.11a-c) are easily verified. The Hamiltonian
$\HH^*_{\rom{II}} \in \II_A $ is taken as
$$
\HH^*_{\rom{II}}:= \HH_{\rom{II}} + {a^2\over \k},   \tag{2.24}
$$
where
$$
\align
\HH_{\rom{II}} =& {1\over{2\k}} \res_{\l=0}\tr(\l^{-1}\NN^2(\l)) \\
   =& {1\over{\k}}((x_2y_1)^2 + tx_2y_1 + {{t^2}\over4} - \k x_1y_2),
\tag{2.25}
\endalign
$$
and the commuting invariant $a\in \II_A $ is
$$
\align
a :=& -{1\over2\k}\res_{\l=0}\tr(\l^{-2}{\NN^2(\l)})\\
   =& x_1 y_1 + x_2 y_2. \tag{2.26}
\endalign
$$

The differential is given by
$$
\d\HH_{\rom{II}}^*=\d\HH_{\rom{II}} +{2a\over \k}\d a ={1\over\k\l}\NN(\l)
-{2a\over \k^2\l^2}\NN(\l),   \tag{2.27}
$$
so, in the notation of eqs.~(1.13), we have
$$
P_{1}(\d\HH_{\rom{II}}^*) = {\l\over 2} \pmatrix 1& 0\\0&-1\endpmatrix
+{1\over \k}\pmatrix -a &-\k y_1 \\ x_2 & a \endpmatrix.
\tag{2.28}
$$
It follows that eq.~(1.15) is satisfied for $\s=1$, and Hamilton's
equations are equivalent to the isomonodromic deformation equation (1.16),
with $\NN(\l)$ defined by (2.23), and
 $$
\DD_t  := {\partial{}\over \partial t}
-{\l\over 2} \pmatrix 1& 0\\0&-1\endpmatrix
+{1\over \k}\pmatrix a &\k y_1 \\ -x_2 & -a \endpmatrix.  \tag{2.29}
$$

   Since $\HH^*_{\rom{II}}$ is given by:
$$
\HH^*_{\rom{II}} = {1\over{\k}}((x_1^2 + x_2^2)y_1^2 + 2x_1x_2y_1y_2 +
 x_2^2y_2^2 + tx_2y_1 + {t^2\over4} - \k x_1y_2),       \tag{2.30}
$$
the mass $m$ in eq.~(0.2) may be taken as
$$
m={\k\over 2}.  \tag{2.31}
$$
The contravariant tensor $g^{ij}$ is then
$$
g^{ij} := \left(\matrix x_1^2 + x_2^2 & x_1x_2\\ x_1x_2 &
x_2^2\endmatrix\right). \tag{2.32a}
$$
and the inverse:
$$
g_{ij} = \left(\matrix {1\over x_2^2} & -{x_1\over x_2^3} \\
 -{x_1\over x_2^3} & {x_1^2 +x_2^2\over x_2^4}\endmatrix\right) \tag{2.32b}
$$
is again Euclidean. Cartesian coordinates for it may be chosen, for $x_2> 0$,
as
$$
\rho = \ln{x_2}, \quad  \mu = {x_1\over x_2}, \tag{2.33a}
$$
with conjugate momenta
$$
\prh = x_1 y_1+x_2 y_2,  \quad \pmu = x_2 y_1. \tag{2.33b}
$$
(This may be viewed as the canonical lift of a map taking the open upper half
of
the $(x_1,x_2)$--plane to the entire $(\rho, \mu)$--plane, the Euclidean metric
determined by the latter.) In terms of the $(\rho, \mu, \prh, \pmu)$
coordinates,
the Hamiltonian is of the form (1.22), with  vector and scalar potentials:
$$
\align
A =& \phantom{-}{1\over2}(\k \mu^2  + t )d\mu  - {1\over 2}\k\mu d\rho
\tag{2.34a}  \\
V =&  -{1\over 4 \k}(\k^2 \mu^4 + 2t \k \mu^2 +\k^2 \mu^2)
+{t^2\over 4}\left(1-{1\over\k}\right), \tag{2.34b}
\endalign
$$
and the constant of motion $a$  is
$$
a = \prh. \tag{2.35}
$$
The magnetic field $B$ is just the constant $-{\k\over 2}$, so the first of the
Lorentz equations (1.23a,b) gives the conserved quantity
$$
{\k \over 2}(\dot{\rho} +\mu) = a. \tag{2.36}
$$
The second is equivalent to Hamilton's equations for the reduced Hamiltonian
$$
\HH^{red}_{II}= {1\over \k} (\pmu^2+ t \pmu - \k a \mu + \k \mu^2\pmu)
+{a^2\over \k} + {t^2\over 4}.  \tag{2.37}
$$
After elimination of the momentum $\pmu$, these give
$$
 \ddot{\mu} =2\mu^3 + {2 t \mu\over \k} + \a,\tag{2.38}
$$
with
$$
\a := {2a +1\over \k},  \tag{2.39}
$$
which, for $\k=2$, is the standard form of $P_{\rom{II}}$.
\medskip \noindent
{\it 2c.\quad  Painlev\'e $\rom{III}$ \hfill}

\nobreak
The moment map for this case is
$$
\align
J_{III}(x_1,x_2,y_1,y_2) =& \NN(\l)  \\
:=&\pmatrix \k t & 0 \\ 0 & - \k t\endpmatrix
 - {1\over{2 \l}} \pmatrix x_1y_1 + x_2y_2 & 2\left(y_1y_2 -
{\k_1\k_2\over x_2^2} +{{\k^2_2 x_1}\over{x_2^3}}\right)\\
-2x_1x_2 & -x_1y_1 - x_2y_2 \endpmatrix \\
&-{\k t\over{2 \l^2}} \pmatrix y_1x_2  & y_1^2 -
 {\k^2_2\over x_2^2}\\ -x_2^2 &
-y_1x_2\endpmatrix \tag{2.40}
\endalign
$$
where $\k\neq 0$, $\k_1$, $\k_2$ are constants and $t$ is the deformation
parameter. This again satisfies the Poisson bracket relations (1.11a-c). The
Hamiltonian $\HH^*_{\rom{III}} \in \II_A $ is taken as
$$
\HH^*_{\rom{III}}:= \HH_{\rom{III}} + {a^2\over 2t},   \tag{2.41}
$$
where
$$
\align
\HH_{\rom{III}}& = {1\over t}\res_{\l=0} \tr(\l \NN^2(\l)) \\
&  = -2 \k^2 ty_1x_2 + {2\over{t}}\left({{(x_1y_1-x_2 y_2)^2}\over 4}
+ \k_1\k_2{x_1\over x_2}- {\k_2}^2{x_1^2 \over x_2^2}\right)
\tag{2.42}
\endalign
$$
and the commuting invariant $a\in \II_A $ is
$$
\align
a :=& -{1\over 2 \k t}\res_{\l=0} \tr(\NN^2(\l)) \\
 =& x_1 y_1 + x_2 y_2. \tag{2.43}
\endalign
$$

The differential is given by
$$
\d\HH_{\rom{III}}^*=\d\HH_{\rom{III}} +{a\over t} \d a =
{2\l\over t}\NN(\l) -{a\over \k t^2}\NN(\l),  \tag{2.44}
$$
so, in the notation of eqs.~(1.13), we have
$$
\align
P_{0}(\d\HH_{\rom{III}}) +P_1\left(\d\left({a^2\over 2t}\right)\right)
 =&  \l\pmatrix \k & 0\\ 0 & -\k\endpmatrix
+{\k\over 2\l} \pmatrix y_1x_2  & y_1^2 -
 {\k^2_2\over x_2^2}\\ -x_2^2 &
-y_1x_2\endpmatrix   \tag{2.45}\\
& -{1\over 2t}\pmatrix3(x_1y_1 + x_2y_2) & 2(y_1y_2 -
{\k_1\k_2\over x_2^2} +{{\k^2_2 x_1}\over{x_2^3}})\\
-2x_1x_2 & -3(x_1y_1 + x_2y_2) \endpmatrix .
\endalign
$$
It follows that eq.~(1.15) is satisfied, with $P_\s(\d\HH)$ replaced by
$P_0(\d\HH_{\rom{III}}) + P_1(\d({a^2\over 2t}))$, and
Hamilton's equations are equivalent to the isomonodromic deformation equation
(1.16),  with $\NN(\l)$ defined by (2.40), and
$$
\align
\DD_t := &{\partial{}\over \partial t } -
P_{0}(\d\HH_{\rom{III}}) - P_1\left(\d\left({a^2\over 2t}\right)\right)\\
 = & {\partial{}\over \partial t } -
\l\pmatrix \k & 0\\ 0 & -\k\endpmatrix
+{1\over 2t}\pmatrix 3(x_1y_1 + x_2y_2) & 2(y_1y_2 -
{{\k_1\k_2}\over{x_2^2}} +{{\k^2_2 x_1}\over{x_2^3}})\\
-2x_1x_2 & -3(x_1y_1 + x_2y_2) \endpmatrix \\
&\phantom{ {\partial{}\over \partial t }}
-{\k\over 2\l} \pmatrix y_1x_2  &
y_1^2 -
 \k^2_2/x_2^2\\ -x_2^2 &
-y_1x_2\endpmatrix.
\tag{2.46}
\endalign
$$

   Since $\HH^*_{\rom{III}}$ is given by:
$$
\HH^*_{\rom{III}} = {1\over t}\left( x_1^2y_1^2 +x_2^2 y_2^2
+2\k_2\k_2 {x_1\over x_2} -2\k_2^2{x_1^2\over x_2^2}\right)
-2\k^2 t y_1 x_2,       \tag{2.47}
$$
either the metric $g_{ij}$ or the mass $m$ in eq.~(0.2) must be chosen as
$t$--dependent.  Taking the mass to be
$$
m={t\over 4},  \tag{2.48}
$$
the contravariant tensor $g^{ij}$ is
$$
g^{ij} := \pmatrix{x_1^2\over 2} &0\\0 &
{x_2^2\over 2}\endpmatrix, \tag{2.49a}
$$
and the inverse:
$$
g_{ij} = \pmatrix {2\over x_1^2} &0\\0 & {2\over
x_2^2}\endpmatrix \tag{2.49b}
$$
is again Euclidean. Cartesian coordinates for it may be chosen, for $x_1,
x_2 > 0$,   as
$$
\rho = \ln{x_1 x_2}, \quad \mu = \ln{x_1\over x_2},    \tag{2.50a}
$$
with conjugate momenta
$$
\prh = {1\over 2}(x_1 y_1 + x_2 y_2),  \quad
\pmu ={1\over 2}(x_1 y_1 - x_2 y_2). \tag{2.50b}
$$
(Again, this may be viewed as the canonical lift of a map taking the first
quadrant of the $(x_1,x_2)$--plane to the entire $(\rho, \mu)$--plane.) In
terms
of these coordinates, the Hamiltonian is of the form (1.22), with  vector and
scalar potentials:
$$
\align
A =& -{\k^2 t^2\over 2}e^{-\mu}d\rho  -{\k^2 t^2\over 2}e^{-\mu}d\mu
\tag{2.51a}  \\
V =&  {2\k_1\k_2\over t}e^\mu -{2\k_2^2\over t}e^{2\mu}
-\k^4t^3e^{-2\mu}, \tag{2.51b}
\endalign
$$
and the constant of motion $a$  is
$$
a = 2\prh. \tag{2.52}
$$

   The magnetic field $B$ is now
$$
B= {1\over 2} \k^2 t^2 e^{-\mu}.  \tag{2.53}
$$
Taking into account the $t$--dependence in the $A_\rho$ component of the
vector potential,
the first of the Lorentz equations (1.23a,b) gives the
conserved quantity
$$
{t\over 2}\left(\dot{\rho}+2\k^2 t e^{-\mu}\right)  = a. \tag{2.54}
$$
The second is equivalent to Hamilton's equations for the reduced Hamiltonian
$$
\HH^{red}_{III}= {2\over t}\left( \pmu -{\k^2t^2e^{-\mu}\over 2}\right)^2
+{2\k_1\k_2 e^{\mu}\over t} - {2\k_2^2 e^{2\mu} \over t}
-a\k^2 t e^{-\mu} -{\k^4 t^3 e^{-2\mu} \over 2} + {a^2\over 2t}.
 \tag{2.55}
$$
After elimination of the momentum $\pmu$, these give
$$
\ddot{\mu} =  -{\dot{\mu}\over t}  - 4(a+2)\k^2 e^{-\mu}
-{8\k_1\k_2 e^{\mu} \over t^2} + {16\k_2^2 e^{2\mu} \over t^2}
-4t^2\k^4 e^{-2\mu}.
\tag{2.56}
$$
Making the change of variable
$$
u:= te^{-\mu}, \tag{2.57}
$$
this becomes
$$
\ddot{u} = {\dot u^2\over u}  -
{\dot{u}\over t} + {1\over t}(\a u^2 + \b) + \g u^3 + {\d \over u}, \tag{2.58}
$$
where
$$
\aligned
\a &:= 4(a+2)\k^2,  \quad   \b := 8\k\k_1\k_2  \\
\g &:= 4 \k^4, \qquad \qquad \d := -16\k^2 \k_2^2,
\endaligned   \tag{2.59}
$$
which is the standard form of $P_{\rom{III}}$.
\medskip \noindent
{\it 2d.\quad  Painlev\'e $\rom{IV}$ \hfill}

\nobreak
The moment map for this case is
$$
\align
J_{IV}(x_1,x_2,y_1,y_2) := \NN(\l) =&\l \pmatrix 1 & 0\\0 &
-1\endpmatrix + \pmatrix 0 & 2 y_1\\ -x_1 & 0\endpmatrix   \\
& + {1\over{2(\l - t)}}\pmatrix -x_2 y_2  & - y_2^2 + {\k^2\over x_2^2}
\\ x_2^2 & x_2 y_2  \endpmatrix \tag{2.60}
\endalign
$$
where $\k$ is a nonzero constant and $t$ is the deformation parameter.
The Poisson bracket relations (1.11a-c) are again easily verified. The
Hamiltonian $\HH^*_{\rom{IV}} \in \II_A$ in this case is taken as
$$
\HH^*_{\rom{IV}}:= \HH_{\rom{IV}} + a^2,   \tag{2.61}
$$
where
$$
\align
\HH_{\rom{IV}}& := {1\over2}\res_{\l = t}\tr(\NN^2(\l)) \\
& = -tx_2y_2 + x_2^2y_1 + {{x_1y_2^2}\over2}- {{\k^2x_1}\over{2x_2^2}},
\tag{2.62}
\endalign
$$
and the commuting invariant $a\in \II_A $ is
$$
\align
a :=& -{1\over 4}\left(\res_{\l = 0}\tr(\l^{-1} \NN^2(\l))
+\res_{\l = t}\tr(\l^{-1} \NN^2(\l))\right) \\
 =&\phantom{-} x_1 y_1 +{1\over 2} x_2 y_2. \tag{2.63}
\endalign
$$

The differential is given by
$$
\d\HH^*_{\rom{IV}}=\d\HH_{\rom{IV}} + 2a \d a =\NN(\l)-{a\over\l}\NN(\l)
\tag{2.64}
$$
so, in the notation of eq.~(1.13), we have
$$
P_{-1}(\d\HH_{\rom{IV}}) + P_1(\d a^2) = - {1\over{2(\l - t)}}\pmatrix -x_2 y_2
 & - y_2^2 +
{\k^2\over x_2^2}\\ x_2^2 & x_2 y_2  \endpmatrix
  - \pmatrix a & 0 \\ 0 & -a\endpmatrix.
\tag{2.65}
$$
Eq.~(1.15) is satisfied with $P_\s(\d\HH)$ replaced by
$P_{-1}(\d\HH_{\rom{IV}}) +
P_1(\d a^2)$, so Hamilton's equations are equivalent to the
isomonodromic deformation equation (1.16),  with $\NN(\l)$ defined by (2.60)
and
$$
\DD_t:= {\partial{}\over \partial t } + \pmatrix a & 0 \\ 0 & -a\endpmatrix
+{1\over{2(\l - t)}}\pmatrix -x_2 y_2  & - y_2^2 + \k^2x_2^{-2}\\ x_2^2 &
x_2 y_2 \endpmatrix
\tag{2.66}
$$

   Since $\HH^*_{\rom{IV}}$ is given by:
$$
\HH^*_{\rom{IV}} = -t x_2 y_2 + x_2^2 y_1 + \left({x_1y_2^2\over2} +
 {x_2^2y_2^2\over 4} + x_1x_2y_1y_2 + x_1^2 y_1^2 -
{\k^2x_1\over 2x_2^2}\right),       \tag{2.67}
$$
we may chose $m=1$ in eq.~(0.2), and the contravariant tensor $g^{ij}$ becomes
$$
g^{ij} = \pmatrix 2 x_1^2 & x_1x_2\\ x_1x_2 & x_1 +
{{x_2^2}\over2}\endpmatrix. \tag{2.68a}
$$
The inverse
$$
g_{ij} = \pmatrix {{1\over 2x_1^2} + {x_2^2\over 4x_1^3}} &
 -{x_2\over 2 x_1^2}\\ -{x_2\over 2x_1^2} & {1\over x_1} \endpmatrix
\tag{2.68b}
$$
is again Euclidean, and Cartesian coordinates for it may be chosen, for
$x_1 > 0$,  as
$$
\rho= {1\over \sqrt{2}}\ln x_1, \quad \mu ={x_2\over \sqrt{x_1}}, \tag{2.69a}
$$
with conjugate momenta
$$
\prh = \sqrt{2}\left(x_1y_1 + {x_2 y_2\over 2}\right),  \quad
\pmu = {\sqrt{x_1}} y_2. \tag{2.69b}
$$
In terms of these coordinates, the Hamiltonian is of the form (1.22), with
vector and scalar potentials:
$$
\align
A =& - (t\mu + {{\mu^3}\over2})d\mu + {{\mu^2}\over \sqrt{2}}d\rho
\tag{2.70a}  \\
V =& -{\k^2\over 2 \mu^2} -
{t^2\mu^2\over 2}-{t\mu^4\over2}-{\mu^6\over8}
 - {\mu^4\over4}, \tag{2.70b}
\endalign
$$
and the constant of motion $a$  is
$$
a = {1\over \sqrt{2}}\prh. \tag{2.71}
$$
The magnetic field $B$ is now
$$
B= \sqrt{2}\mu,   \tag{2.72}
$$
so the first of the Lorentz equations (1.23a,b) just gives the
conserved quantity
$$
{\dot{\rho}\over \sqrt{2}}-{\mu^2\over 2}  = a, \tag{2.73}
$$
while the second is equivalent to Hamilton's equations for the reduced
Hamiltonian
$$
\HH^{red}_{IV}=
 {1\over2}\pmu^2 +a^2  - \left(t\mu +{\mu^3\over 2}\right)\pmu +
a \mu^2 - {\k^2 \over 2 \mu^2 }. \tag{2.74}
$$
After elimination of the momentum $\pmu$, these give
$$
 \ddot{\mu} = -\a\mu +{\b \over 2 \mu^3}+ 2 t \mu^3 +t^2\mu
 +{3\over 4}\mu^5     \tag{2.75}
$$
with
$$
\a := 2a + 1, \quad \b := -2\k^2 . \tag{2.76}
$$
In terms of the new variable
$$
u:= \mu^2, \tag{2.77}
$$
this becomes
$$
\ddot{u} = {\dot{u}^2\over 2u}+{3u^3\over 2}+ 4t u^2+2(t^2-\a)u +{\b \over u},
\tag{2.78}
$$
which is the standard form of $P_{\rom{IV}}$.
\medskip \noindent
{\it 2e.\quad  Painlev\'e $\rom{V}$ \hfill}

\nobreak
The moment map (1.7) for this case is defined as
$$
\align
J(x_1,x_2,y_1,y_2) := \NN(\l) =& \pmatrix t & 0\\0 & -t\endpmatrix +
{1\over{2 \l}} \pmatrix -x_1 y_1  & -y_1^2 + {\k_1^2 \over x_1^2}\\
 x_1^2 & x_1 y_1 \endpmatrix \\
& + {1\over{2(\l -1)}} \pmatrix
-x_2 y_2  & -y_2^2 + {\k_2^2\over x_2^2}\\ x_2^2 & x_2 y_2 \endpmatrix
\tag{2.79}
\endalign
$$
where $\k_1, \k_2$ are arbitrary constants and $t$ is the deformation
parameter. Again, the Poisson bracket relations (1.11a-c) are easily verified.
The Hamiltonian $\HH^*_{\rom{V}} \in \II_A$  is taken (cf.~\cite{H, HW2}) to be
$$
\align
\HH_{\rom{V}}^*=& \HH_{\rom{V}} -{a^2\over 4t}\\
=&-{1\over4t}(x_1^2+x_2^2)(y_1^2+y_2^2)+
{1\over 4t}\left(\k_1^2{x_2^2\over x_1^2}+\k_2^2{x_1^2\over x_2^2}\right)
-x_2 y_2 +{\k_1^2 \over 4t}
 \tag{2.80}
\endalign
$$
where
$$
\align
\HH_{\rom{V}}:=& {1\over 2t}\left(\res_{\l=0}(\l\NN^2(\l))
+\res_{\l=1}(\l\NN^2(\l))\right) \\
=&-{1\over4t}(x_1 y_2-x_2 y_1)^2 +
{1\over 4t}\left(\k_1^2{x_2^2\over x_1^2}+\k_2^2{x_1^2\over x_2^2}\right)
-x_2 y_2 +{\k_1^2 \over 4t}
\tag{2.81}
\endalign
$$
and the commuting invariant $a\in \II_A $ is
$$
\align
a :=& -{1\over 2t}\left(\res_{\l = 0}\tr(\NN^2(\l))+
\res_{\l = 1}\tr(\NN^2(\l))\right)  \\
 =&\phantom{-{1\over 2t}(} x_1 y_1 + x_2 y_2.
\tag{2.82}
\endalign
$$
The differential is given by
$$
\d\HH_{\rom{V}}^*= \d\HH_{\rom{V}}-{a\over 2t}=
{\l\over t}\NN(\l) +{a\over 2t^2}\NN(\l) \tag{2.83}
$$
so, in the notation of eqs.~(1.13), we have
$$
P_{1}(\d\HH_{\rom{V}}^*) = \l\pmatrix 1 & 0 \\ 0 & -1 \endpmatrix
  + {1\over 2t} \pmatrix 0 & -y_1^2 - y_2^2 + {\k_1^2\over x_1^2} +
{\k_2^2\over x_2^2}\\ x_1^2 + x_2^2 & 0 \endpmatrix.
\tag{2.84}
$$
Eq.~(1.15) is satisfied for $\s=1$, so Hamilton's
equations are equivalent to the isomonodromic deformation equation (1.16),
with $\NN(\l)$ defined by (2.79) and
$$
\DD_t :={\partial\over\partial t}-\pmatrix \l&0 \\0&-\l\endpmatrix-
{1\over 2t} \pmatrix 0&-y_1^2-y_2^2+{\k_1^2\over x_1^2}+{\k_2^2\over x_2^2}\\
x_1^2+x_2^2 & 0 \endpmatrix. \tag{2.85}
$$

  From the form (2.80) of $\HH^*_{\rom{V}}$  we may choose
$$
m=-2t,  \tag{2.86}
$$
and the contravariant tensor $g^{ij}$ becomes
$$
g^{ij} = \pmatrix x_1^2 + x_2^2 & 0\\
 0 & x_1^2 + x_2^2\endpmatrix, \tag{2.87a}
$$
with inverse
$$
g_{ij} =  \pmatrix {1\over x_1^2 + x_2^2} & 0\\
 0 & {1\over{x_1^2 + x_2^2}}\endpmatrix,
\tag{2.87b}
$$
which is again Euclidean. Cartesian coordinates $(\rho, \mu)$ for it may be
chosen  as
$$
\rho = \ln \sqrt{x_1^2 + x_2^2},  \quad x_1= e^\rho \cos \mu,
\quad x_2 = e^\rho \sin\mu, \tag{2.88a}
$$
with conjugate momenta
$$
\prh = x_1 y_1 + x_2 y_2, \quad \pmu= x_1 y_2 - x_2 y_1.    \tag{2.88b}
$$
(Note that, whereas these appear like polar coordinates, metrically, they
define a map from $\bold{R}^2-(0,0)$ to a flat cylinder.) In terms of these
coordinates, the Hamiltonian is of the form (1.22), with  vector and scalar
potentials:
$$
\align
A =& 2t(\sin{\mu}\cos{\mu} d\mu + \sin^2{\mu}d\rho) \tag{2.89a}  \\
V =& {1\over{4t}}(\k^2_1 {\tan \mu}^2 + \k^2_2 {\cot \mu}^2 )
+t\sin^2{\mu} -{\k_1^2 + \k_2^2\over 8t},
\tag{2.89b}
\endalign
$$
and the constant of motion $a$  is
$$
a = \prh. \tag{2.90}
$$\
The magnetic field $B$ is now
$$
B= 2t\sin{2\mu},   \tag{2.91}
$$
so the first of the Lorentz equations (1.23a,b) just gives the
conserved quantity
$$
-2t \dot{\rho} -2t\sin^2\mu= a \tag{2.92}
$$
and the second is equivalent to Hamilton's equations for the reduced
Hamiltonian
$$
\HH^*_{\rom{V}} = -{\pmu^2\over 4t}+{\k^2_1\tan^2\mu + \k^2_2\cot^2\mu\over 4t}
-
a\sin^2\mu - \pmu\sin{\mu}\cos{\mu} +{\k_1^2-a^2\over 4t}. \tag{2.93}
$$
In terms of the new variable
$$
u:= -\cot^2{\mu} \tag{2.94a}
$$
and its conjugate momentum
$$
p_u := {\sin^2{\mu}\pmu\over 2\cot{\mu}}, \tag{2.94b}
$$
this becomes
$$
\HH^*_{\rom{V}} = {u(1-u)^2p_u^2\over t} + 2u p_u - {a\over 1- u} -
- {\k_1^2\over 4t u} - {k_2^2 u\over 4t}
+{\k_1^2 - a^2\over 4t},  \tag{2.95}
$$
and Hamilton's equations, after elimination of the momentum $p_u$ are
$$
\ddot{u} = \left({1\over 2u}+{1\over u-1}\right) \dot{u}^2 -{1\over t}\dot{u}
+{(\a u^2 +\b)(u-1)^2\over t^2 u} +{\g u\over t} +{\d u(u+1)\over u-1},
\tag{2.96}
$$
where
$$
\aligned
\a :=& {\k_2^2\over 2} , \quad  \quad \b :=-{\k_1^2 \over2}\\
\gamma :=& 2a+2, \qquad \quad \d := -2,
\endaligned
 \tag{2.97}
$$
which is the standard form of $P_{\rom{V}}$.
\bigskip
\subheading{3.  Discussion}

\nobreak
   We have seen that both the Hamiltonian structure and the isomonodromic
deformation property of the Painlev\'e transcendents arise naturally within
the framework of time dependent Poisson embeddings of a $4$--dimensional
phase space into the loop algebra $\wt{\frak{sl}}(2)_R$. The Hamiltonian
structure
of the Painlev\'e transcendents can, of course, be simply interpreted in the
reduced system as due to time dependent scalar potentials in
$1$--dimension. But this interpretation does not  provide a framework from
which
to derive the Hamiltonian structure, nor does it lead in any intrinsic way to
the
corresponding isomonodromic deformation equations. The $2$--dimensional
electromagnetic models derived here, however, follow naturally from the
$2\times
2$ rational matrix form provided by the $\wt{\frak{sl}}(2)_R$ loop algebra in
the classical $R$--matrix approach. The pair consisting of the Hamiltonian
$\HH_A$
and the symmetry generator $a$ required for reduction to a $1$--dimensional
system are both seen in this formulation as generators of the ring $\II_A$ of
spectral invariants on a Poisson subspace $\frak{g}_A\ss\wt{\frak{sl}}^*(2)_R$.
The isomonodromic deformation equations follow by modifying the isospectral
equations resulting from the $R$--matrix Poisson structure by taking into
account
the $t$--dependent moment map $J_A$.

   The loop algebra approach to isomonodromic deformation equations for matrix
differential operators of arbitrary rank having regular singularities
in the finite $\l$--plane was developed in \cite{H}. The examples of
the Painlev\'e transcendents $P_{\rom{I}}$--$P_{\rom{V}}$ show that this theory
may be extended, within the classical $R$--matrix setting, to the case of
irregular singular points  both at finite and infinite values of $\l$. The
general case, with arbitrary rational $\NN(\l)$ of any rank \cite{JMU, JM}, is
also likely to be amenable to a Hamiltonian formulation within the $R$--matrix
framework using the $t$--dependent version of the general moment map
construction
of refs. \cite{AHP, AHH, HW1}.

\bigskip \bigskip
 \centerline {\bf References}
\bigskip {\smaller
\item{\bf [AHH]} Adams, M.R., Harnad, J. and Hurtubise, J.,
``Dual Moment Maps into Loop Algebras'', {\sl Lett. Math. Phys.} {\bf 20}
(1990),
299--308.
\item{\bf [AHP]} Adams, M.R., Harnad, J. and Previato, E.,
``Isospectral Hamiltonian Flows in Finite and Infinite Dimensions I.
Generalized Moser Systems and Moment Maps into Loop Algebras'',
{\sl Commun. Math. Phys.} {\bf 117} (1988), 451--500.
\item{\bf [FT]} Faddeev, L.D\. and Takhtajan, L\.A\.,
 {\it Hamiltonian Methods in the Theory of Solitons},\break
 Springer--Verlag, Heidelberg (1987).
\item{\bf [H]} Harnad, J., ``Dual Isomonodromic Deformations and Moment Maps
into  Loop Algebras'', {\sl Commun. Math. Phys} (1994, in press).
\item{\bf [HW1]} Harnad, J.  and Wisse, M.--A., ``Moment Maps to Loop Algebras,
Classical $R$--Matrix  and Integrable Systems'', in: {\sl Quantum Groups,
Integrable Models  and Statistical Systems}  (Proceedings of the 1992 NSERC-CAP
Summer Institute in Theoretical Physics, Kingston, Canada, July 1992), ed.
J.~Letourneux and  L.~Vinet, World Scientific, Singapore (1993).
\item{\bf [HW2]}  Harnad, J.  and Wisse, M.--A., `` Loop Algebra Moment Maps
and
Hamiltonian Models for the Painlev\'e Transcendents'', in: Mechanics Day
Workshop Proceedings,  	ed. P.S. Krishnaprasad, T. Ratiu and W.F. Shadwick,
(AMS - Fields Institute Commun., 1994).
 \item{\bf [I]} Ince, E.L.,
``Ordinary Differential Equations'', Dover New York 1956.
 \item{\bf[JMU]} Jimbo, M., Miwa, T., Ueno, K., ``Monodromy Preserving
Deformation of Linear Ordinary Differential Equations with Rational
Coeefficients I.'', {\it Physica} {\bf 2D}, 306-352 (1981).
 \item{\bf[JM]} Jimbo, M., Miwa, T., ``Monodromy Preserving
Deformation of Linear Ordinary Differential Equations with Rational
Coeefficients II, III'', {\it Physica} {\bf 2D}, 407-448 (1981); {\it ibid.},
{\bf 4D}, 26-46 (1981).
\item{\bf [Ok]} Okamoto, K., ``On the $\tau$-function of the Painlev\'e
equations'',  {\sl Physica  D} {\bf 2} (1981) 525--535;
``The Painlev\'e equations and the Dynkin Diagrams'',  in: {\sl Painlev\'e
Transcendents. Their Asymptotics and Physical Applications}, ed. P.~Winternitz
and D.~Levi, Plenum Press, N.Y., {\sl NATO ASI Series B} Vol. {\bf 278} (1992)
299-313.
\item{\bf [ST]} Semenov-Tian-Shansky, M.A., ``What is a classical
$R$-matrix'', {\sl Funct. Anal. Appl.} {\bf 17} (1983) 259--272. \medskip }
\vfill \eject
\enddocument